\documentclass[12pt,preprint,nofootinbib,noshowkeys,noshowpacs,longbibliography,aps,prd,
superscriptaddress,tightenlines,floatfix]{revtex4-2}
\usepackage{amsmath,amsfonts,amsthm,amssymb}
\usepackage{graphics,graphicx}
\usepackage{color}
\definecolor{darkblue}{RGB}{0,0,196}
\definecolor{darkgreen}{RGB}{0,120,0}

\def\beq{\begin{equation}}
\def\eeq{\end{equation}}
\def\st{\begin{equation}}
\def\stp{\end{equation}}
\def\ba{\begin{eqnarray}}
\def\ea{\end{eqnarray}}

\usepackage[colorlinks=true,linktocpage=true,linkcolor=darkblue,citecolor=red,urlcolor=darkblue,hyperfootnotes=true]{hyperref}

\begin{document}
\preprint{}
 
    \title{Field redefinition and its impact in relativistic hydrodynamics}    
    \author{Sayantani Bhattacharyya}
    \email{sbhatta5@ed.ac.uk}
    \affiliation{School of Mathematics, University of Edinburgh, Peter Guthrie Tait Road, Edinburgh EH9 3FD, United Kingdom}
    \affiliation{School of Physical Sciences, National Institute of Science Education and Research, An OCC of Homi Bhabha National Institute, Jatni-752050, India}
    \author{Sukanya Mitra}
    \email{sukanya.mitra@niser.ac.in}
    \affiliation{School of Physical Sciences, National Institute of Science Education and Research, An OCC of Homi Bhabha National Institute, Jatni-752050, India}
    \author{Shuvayu Roy}
    \email{shuvayu.roy@iitgn.ac.in}
    \affiliation{School of Physical Sciences, National Institute of Science Education and Research, An OCC of Homi Bhabha National Institute, Jatni-752050, India}
    \affiliation{Indian Institute of Technology, Gandhinagar, Gujarat 382355, India}
\author{Rajeev Singh}
    \email{rajeev.singh@e-uvt.ro}
    \affiliation{School of Physical Sciences, National Institute of Science Education and Research, An OCC of Homi Bhabha National Institute, Jatni-752050, India}
    \affiliation{Department of Physics, West University of Timisoara, Bd.~Vasile P\^arvan 4, Timisoara 300223, Romania}
    \affiliation{Center for Nuclear Theory, Department of Physics and Astronomy, Stony Brook University, Stony Brook, New York, 11794-3800, USA}
	\date{\today} 
	\bigskip
\begin{abstract}
In this paper, we explore the impact of field redefinition on the spectrum of linearized perturbations in relativistic hydrodynamics. We observe that the spectrum of hydrodynamics modes is never affected by the local field redefinition, however, the spectrum of the non-hydrodynamic modes is affected. Through an appropriate all-order redefinition, non-hydrodynamic modes can be eliminated, leading to a new frame where the spectrum contains only hydrodynamic modes. We also observe that the resulting stress-energy tensor may have an infinite series in momentum space, with a convergence radius linked to the eliminated non-hydrodynamic mode. In certain special cases, higher-order terms in the stress-energy tensor under field redefinition may cancel, indicating that non-hydrodynamic modes are mere artefacts of the fluid variable choice and hold no physical significance, even if they appear to violate physical constraints. Using a special toy example, we find a criterion to distinguish between physical and unphysical non-hydrodynamic modes.
\end{abstract}
     
\date{\today}
	
\maketitle
\newpage

\section{Introduction}
\label{sec:intro}
%
In the formulation of relativistic fluid dynamics, it is crucial for any acceptable model to satisfy important physical constraints such as causality, stability etc. These constraints are typically tested by analyzing the behavior of linearized perturbations around an equilibrium background. When the background equilibrium exhibits translational symmetry, the analysis is performed using Fourier modes, and the spectrum of these modes is scrutinized to ensure that the model adheres to the required physical constraints.
To ensure that the criteria for validating a fluid model are robust and observer-independent, the analysis must respect symmetries inherent in the system. For instance, if the background fluid possesses rotational invariance, then all spatial coordinates related by global rotations must be treated equivalently. This means that the observer can choose any set of coordinates, and the constraints derived from the analysis should remain consistent regardless of this choice.
Rotational invariance is inherently incorporated into the perturbation analysis, which ensures that this symmetry is explicitly reflected in the spectrum of Fourier modes and the resulting constraints. As a result, the physical validity of the fluid model remains unaffected by the specific choice of coordinates made by the observer, thus preserving the model's robustness and universality~\cite{Hiscock:1985zz,Geroch:1990bw,Geroch:1995bx,Gavassino:2021kjm,Arnold:2014jva,Lehner:2017yes}.

However, such manifest invariance is not guaranteed for every gauge freedom that a model might have. When invariance is not manifest, it becomes crucial to disentangle the artefacts of choice from the real physical spectrum so that constraints arising from stability or causality can be meaningfully tested.
For example, in relativistic hydrodynamics, ambiguity arises in the definition of fluid variables like velocity and temperature as soon as we move beyond the perfect fluid scenario. In thermodynamic equilibrium, these variables have natural and unambiguous definitions in terms of the spatially and temporally uniform thermodynamic potentials. However, outside equilibrium, they can be redefined by adding arbitrary corrections that are nonzero only if the fluid variables exhibit spacetime variations. These ambiguities are typically resolved by imposing external constraints on the relativistic stress tensor or current, similar to the method of gauge fixing in gauge theory. Common choices include the `Landau frame' or `Eckart frame'~\cite{Israel:1976tn,Israel:1979wp,Bemfica:2019cop}. 

Nevertheless, one is always free to choose constraints other than these two options, or even proceed without fixing these field ambiguities, as is done in the recent BDNK (Bemfica-Disconzi-Noronha-Kovtun) formalism~\cite{Bemfica:2017wps,Kovtun:2019hdm}. It turns out that the spectrum of linearized perturbations is non-trivially modified when the fluid fields are redefined. The BDNK formalism, in particular, keeps the fluid frame unfixed to determine the frame where first-order fluid dynamics (stress tensor with the first sub-leading corrections beyond the ideal fluid scenario) maintains causality and stability. This approach contrasts with the known issues of causality and stability in the usual Landau or Eckart frame~\cite{Denicol:2008ha,Floerchinger:2017cii,Biswas:2020rps,Liu:2020ksx,Hoult:2023clg,Sammet:2023bfo,Domingues:2024pom}.

In summary, while rotational and translational symmetries are straightforwardly handled in the analysis, gauge freedoms in the definition of fluid variables introduce additional complexities. Properly accounting for these complexities is essential for ensuring that the physical constraints derived from the analysis of linearized perturbations genuinely reflect the underlying physics, free from artefacts introduced by arbitrary choices in the definition of fluid variables~\cite{Noronha:2021syv,Mitra:2023ipl,Bhattacharyya:2023srn,Bhattacharyya:2024tfj,Mitra:2024yei,Xie:2023gbo}.

It is not surprising that the spectrum changes under field redefinition because the perturbations of the redefined fields are inherently different physical quantities from the perturbations of the original fields~\cite{Grozdanov:2019uhi,Dore:2021xqq,Armas:2022wvb,Bea:2023rru,Hoult:2024cyx}. These redefined fields satisfy different linear differential equations, resulting in a different spectrum. However, whether a particular model of relativistic fluid dynamics is causal should not depend on the choice of variables used to describe the model.

In Ref.~\cite{Bhattacharyya:2024jxm}, we studied how field redefinition influences the spectrum of linearized perturbations in relativistic fluid dynamics and how non-hydrodynamic modes could be removed using an appropriate all-order field redefinition. In this paper, we provide details of the derivation and extended our analysis.
We have observed that
\begin{enumerate}
\item The spectrum of hydrodynamic modes (modes whose frequency vanishes in the limit of vanishing spatial momentum) is never affected by local field redefinition.
\item The spectrum of non-hydrodynamic modes (modes whose frequency tends to a finite nonzero value in the limit of vanishing momenta) is affected by field redefinition.
\item All non-hydrodynamic modes can be eliminated from the spectrum through an appropriate, generically all-order, field redefinition, see related investigations~\cite{Grozdanov:2018fic}. After applying such an all-order field redefinition, the spectrum of the linearized perturbations will be simplified, containing only the hydrodynamic modes of the original fluid model. We shall refer to this new fluid frame as the ``hydro frame''. Even if we begin with a stress tensor containing a finite number of terms (as in BDNK theory), transforming to the hydro frame will typically result in a stress tensor with an infinite number of terms, involving arbitrarily high-order derivatives of the fluid variables.
We know that this modified stress-tensor may not be suitable for numerical implementation but our motive is to understand the nature of non-hydrodynamic modes.
\item If we re-express the infinite-order stress tensor of the hydro frame in momentum space (in terms of the frequency $\omega$ and spatial momentum $k$ of the linearized perturbation), the resulting infinite series in the $\{\omega, k\}$ plane will have a radius of convergence determined by the non-hydrodynamic mode whose removal via field redefinition generated the infinite series in the stress tensor. This illustrates how the non-hydrodynamic mode can control the validity of the hydrodynamic expansion.
 \item In some very special cases, even after implementing the infinite-order field redefinition, the stress tensor in the hydro frame may still contain a finite number of terms, with higher-order terms precisely canceling each other and thus always convergent. In such scenarios, the information about the non-hydrodynamic mode can simply be erased by field redefinition. These non-hydrodynamic modes are merely artefacts of the choice of fluid variables and lack any physical significance. More importantly, if such a mode violates the criteria for causality, stability, etc., it does not imply that the particular fluid model in question is unphysical.
 \item If the dispersion polynomial do not cleanly factorizes between hydrodynamic and non-hydrodynamic modes, then those non-hydrodynamic modes are not frame artefacts but physical.
\end{enumerate}
The structure of this paper is as follows: In Section~\ref{sec:setup}, we provide a detailed setup for our study. Section~\ref{sec:fluid} investigates the impact of field redefinition on the dispersion relation in the context of relativistic fluid dynamics. In Section~\ref{sec:framerem}, we present an algorithm to remove non-hydrodynamic modes, followed by its application to the BDNK framework in Section~\ref{sec:BDNK}. Section~\ref{sec:artificial} examines artificial non-hydrodynamic modes and demonstrates their elimination using a toy example. In Section~\ref{sec:special_correction}, we use a specific second-order correction to BDNK to establish another criterion for identifying physical non-hydrodynamic modes. Finally, we conclude our findings in Section~\ref{sec:conclude}.
\section{The setup}
\label{sec:setup}
In this section we would like to investigate how field redefinition of fluid variables affects the dispersion polynomial of any linearization problem.

Suppose we have a set of variables $\{\Phi_i\}$ (for a fluid, these may be velocity, temperature and other conserved charges) satisfying a set of nonlinear coupled PDEs collectively denoted as  ${\cal E}(\{\Phi_i\})=0$ (for fluid ${\cal E}$ consists of PDEs resulting from stress tensor and charge current conservation). Further, we assume $\{\Phi_i\} = \{\bar \Phi_i\}$ is an exact solution of ${\cal E}$ that is invariant under space-time translation and spatial rotation. To compute the spectrum we first linearize the equation ${\cal E}$ around $\{\bar\Phi_i\}$
\begin{equation}\label{lin1}
\begin{split}
\text{Define}~~~~& \Phi_i^s=\bar \Phi_i + \epsilon~ \delta\Phi_i(\omega,k) e^{-i\omega t + i\vec k\cdot\vec x},~~~k = \sqrt{\vec k\cdot \vec k},~~\epsilon<<1\,,\\
&{\cal E}(\{\Phi_i= \Phi_i^s\})=0~\Rightarrow~\sum_j M_{ij}(\bar \Phi,\omega,k)~\delta\Phi_j=0\,.
\end{split}
\end{equation}
The spectrum is determined from the zeroes of the determinant of the linearization matrix $M_{ij}$. If the equation ${\cal E}$ has finite number of derivatives,  $Det[M]$ will be a finite polynomial in $\omega$ and $k$ and therefore, will have finite number of zeroes.

Now suppose we are implementing a field redefinition $\Phi_i\rightarrow\Psi_i = \Phi_i + \Delta \Phi_i$. Here, the shift field $\Delta\Phi$ is a nonlinear function of $\Phi_i$ and its derivatives. To relate to the situation in fluid dynamics,  we further  assume that, in equilibrium, both $\Phi_i$ and $\Psi_i$ agree i.e., $\Delta\Phi_i$ vanishes when evaluated on $\Phi_i = \bar \Phi_i$. Under such field redefinition the equation of motion transforms as ${\cal E}(\Phi)\rightarrow\tilde{\cal E}(\Psi)$.  Though the set of PDEs $\tilde{\cal E}$ could be completely determined once we know ${\cal E}$ and the field transformation $\Delta\Phi$,  it will have a very  different structure than that of ${\cal E}$.  We could linearize $\tilde{\cal E}$ to get the spectrum exactly the way we did  for ${\cal E}$ in equation \eqref{lin1} leading to a different linearization matrix $\tilde M_{ij}$. Our goal is, then, to find the relation between the two matrices $M_{ij}$ and $\tilde M_{ij}$
\begin{equation}\label{lin2}
\begin{split}
\text{Define}~~~~&\Psi_i^s = \bar \Phi_i + \epsilon~ \delta\Psi_i(\omega,k) e^{-i\omega t + i\vec k\cdot\vec x}\,,~~~k = \sqrt{\vec k\cdot \vec k},~~\epsilon<<1\,,\\
&\tilde{\cal E}(\{\Psi_i=\Psi_i^s\})=0~\Rightarrow~\sum_j \tilde M_{ij}(\bar \Phi,\omega,k)~\delta\Psi_j=0\,.
\end{split}
\end{equation}
Since $\Delta\Phi$  vanishes when $\Phi =\bar\Phi$, it must be of order ${\cal O}(\epsilon)$  or higher when evaluated on field configuration as described in equation \eqref{lin1}. In other words, generically $\Delta\Phi$ could be expressed in terms of the following matrix equation\footnote{We could design a field redefinition such that the order ${\cal O}(\epsilon)$ term in $\Delta\Phi$ also vanishes. But such redefinitions will not have any impact on the spectrum of the linearized perturbation and, therefore, are not relevant for our purpose.}
\begin{equation}\label{shiftlin}
\begin{split}
\Delta\Phi_i (\Phi^s)= \epsilon\sum_j S_{ij}(\bar\Phi,\omega,k)~\delta\Phi_j(\omega,k) e^{-i\omega t + i\vec k\cdot\vec x}+ {\cal O}(\epsilon^2)\,.
\end{split}
\end{equation}
Now suppose under field redefinition $\Phi_i^s\rightarrow \Psi_i^s$
\begin{equation}\label{translin2}
\begin{split}
\Psi_i^s &= \bar \Phi_i + \epsilon~ \delta\Psi_i(\omega,k) e^{-i\omega t + i\vec k\cdot\vec x}\,, \\
&= \Phi_i^s + \Delta\Phi_i (\Phi^s)\,, \\
& = \bar \Phi_i + \epsilon~ \delta\Phi_i(\omega,k) e^{-i\omega t + i\vec k\cdot\vec x}+ \epsilon\sum_j S_{ij}(\bar\Phi,\omega,k)~\delta\Phi_j(\omega,k) e^{-i\omega t + i\vec k\cdot\vec x}+ {\cal O}(\epsilon^2)\,,\\
& = \bar \Phi_i + \epsilon\sum_j\left[\delta_{ij} + S_{ij}(\bar\Phi,\omega,k)\right]~\delta\Phi_j(\omega,k) e^{-i\omega t + i\vec k\cdot\vec x}+ {\cal O}(\epsilon^2)\,,\\
\end{split}
\end{equation}
it follows that
\begin{equation}
    \delta\Psi_i(\bar\Phi,\omega,k) = \sum_j\left[\delta_{ij} + S_{ij}(\bar\Phi,\omega,k)\right]\delta\Phi_j(\bar\Phi,\omega,k)\,.
\end{equation}
Substituting the above relation in equation \eqref{lin2} and then comparing with equation \eqref{lin1} we find~\cite{Bhattacharyya:2024jxm}
\begin{equation}\label{finclude}
\begin{split}
&\sum_{jk} \tilde M_{ij} \left(\delta_{jk} + S_{jk}\right)\delta\Phi_k=0\,,\\
\Rightarrow ~& M_{ik} = \sum_{j} \tilde M_{ij} \left(\delta_{jk} + S_{jk}\right)~~\Rightarrow~~Det[M] = Det[\tilde M] ~~Det[ {\bf 1} + S]\,.
\end{split}
\end{equation}
Clearly, the zeroes of $Det[ {\bf 1} + S]$ will look like the new modes in the `$\Phi$ frame' which are not present in the `$\Psi$ frame'. These new zero modes are nothing but those form of perturbations in the `$\Phi$ frame', where the new pieces generated  by the linearized field redefinitions are precisely cancelled by the linearized fluctuations.  From the perspective of `$\Psi$ frame' these new modes correspond to no fluctuations at all. Clearly these new modes are artefacts of  frame transformation and therefore should not have any physical basis. More importantly, depending on our choice of field redefinitions, such artificial modes might look unstable or acausal even when the theory in `$\Psi$ frame' is perfectly sensible. But if we are in the `$\Phi$ frame',  a priori we do not have a way to distinguish such artefacts from the physical ones.
In the subsequent sections we would like to address this particular issue in the context of relativistic fluid dynamics.
\section{Specializing to Fluid Dynamics}\label{sec:fluid}
In the previous section, we explored how field redefinitions can impact the dispersion relation for any system of equations. In this section, we focus specifically on fluid variables and determine the most general form of the frame transformation matrix $S$ introduced earlier (see equation \eqref{shiftlin}). To keep the discussion straightforward, we will limit our analysis to uncharged fluids, where the fluid velocity and temperature are the only variables, and the equation of motion is governed by the conservation of the stress tensor.

Let us denote the velocity and the temperature in two different frames `Frame-1' and `Frame-2' as $\{\hat u^\mu,\hat T\}$ and $\{u^\mu,T\}$, respectively; shift functions $\Delta u^\mu$ and $\Delta T$ are defined as
\begin{equation}\label{fluidframe}
u^\mu = \hat u^\mu +\Delta u^\mu (\hat u,\hat T)\,, \qquad T = \hat T + \Delta T(\hat u, \hat T)\,.
\end{equation}
In a rotationally and translationally invariant equilibrium, we expect the fluid variables in two different frames to coincide. As a result, every term in $\Delta u^\mu$ and $\Delta T$ must involve at least one space-time derivative, ensuring that the shift variables are only non-zero for spatially and/or temporally non-uniform fluid profiles. Moreover, we are interested only in those terms that, when evaluated on fluid profiles consisting of equilibrium plus small fluctuations, contribute at linear order in the fluctuation amplitude. This applies to all terms with a single derivative, while for terms involving multiple derivatives, all derivatives must act on a single fluid variable. To clarify this point, let us examine the possible `two-derivative' terms that could appear in $\Delta u^\mu$
\begin{enumerate}
 \item $T_1^\mu = (u^\alpha u^\beta\partial_\alpha\partial_\beta)u^\mu$\,,
 \item $T_2^\mu = (u^\alpha\partial_\alpha u^\beta)\partial_\beta u^\mu$\,.
 \end{enumerate}
In $T_1^\mu$ both the derivatives act on the same $u^\mu$ and, therefore, it will contribute at the linear order in the amplitude fluctuation. The other term $T_2^\mu$ also has two derivatives, but they act on two different $u^\mu$. Each of the `factors'  $(u^\alpha\partial_\alpha u^\beta)$ and $\partial_\beta u^\mu$ will be nonzero at linear order and therefore their product must be quadratic or higher in the fluctuation amplitude. Terms of the form of $T_2^\mu$ will certainly be there in $\Delta u^\mu$ but will have no impact on the spectrum of linearized fluctuation. 

After considering the aspects mentioned above, the most general form of $\Delta u^\mu$ and $\Delta T$ that could contribute to the spectrum of  small fluctuation is the following~\cite{Bhattacharyya:2024jxm}
\begin{equation}\label{motstg}
\begin{split}
\Delta u^\mu &= F_u[(\hat u\cdot\partial),\hat P^{\alpha\beta}\partial_\alpha\partial_\beta] (\hat u\cdot\partial) \hat u^\mu + F_T[(\hat u\cdot\partial),\hat P^{\alpha\beta}\partial_\alpha\partial_\beta] \left( \frac{\hat P^{\mu\alpha}\partial_\alpha \hat T}{\hat T}\right)\\
&+ R_u[\hat P^{\alpha\beta}\partial_\alpha\partial_\beta] \left(\hat P^{\mu\theta}\hat P^{\alpha\beta} \partial_\alpha\partial_\beta \hat u_\theta\right)\,,\\
{\Delta T\over\hat  T}&= G_u[(\hat u\cdot\partial),\hat P^{\alpha\beta}\partial_\alpha\partial_\beta] \left(\partial\cdot \hat u\right) + G_T[(\hat u\cdot\partial),\hat P^{\alpha\beta}\partial_\alpha\partial_\beta] \left(\frac{\hat  u^\alpha\partial_\alpha\hat T}{\hat  T}\right)\\
&+ R_T[\hat P^{\alpha\beta}\partial_\alpha\partial_\beta]\,\left(\frac{\hat P^{\alpha\beta} \partial_\alpha\partial_\beta \hat T}{\hat T}\right)\,,
\end{split}
\end{equation}
where $F_u$, $F_T$, $G_u$, $G_T$ are all linear differential operators of the form
\begin{equation}\label{definc}
\begin{split}
F_u &\equiv \sum_{m,n} f_{m,n}^u \left[\hat u\cdot\partial\right]^m \left[\hat P^{\alpha\beta}\partial_\alpha\partial_\beta\right]^n\,, \quad
F_T \equiv \sum_{m,n} f_{m,n}^T \left[\hat u\cdot\partial\right]^m \left[\hat P^{\alpha\beta}\partial_\alpha\partial_\beta\right]^n\,,\\
G_u &\equiv \sum_{m,n} g^u_{m,n} \left[\hat u\cdot\partial\right]^m \left[\hat P^{\alpha\beta}\partial_\alpha\partial_\beta\right]^n\,, \quad G_T \equiv \sum_{m,n} g^T_{m,n} \left[\hat u\cdot\partial\right]^m \left[\hat P^{\alpha\beta}\partial_\alpha\partial_\beta\right]^n\,,
\end{split}
\end{equation}
whereas,
$R_T$ and $R_u$ have a little special structure in the sense that they do not depend on the differential operator $(\hat u\cdot\partial)$. They are introduced to capture those field redefinitions that do not vanish in the limit of vanishing frequency ($\omega \rightarrow 0$). They also admit an expansion in the small $k$ limit as
\begin{equation}
R_T =  \sum_{m} r^T_{m} \left[\hat P^{\alpha\beta}\partial_\alpha\partial_\beta\right]^m \,,\qquad
R_u =  \sum_{m} r^u_m  \left[\hat P^{\alpha\beta}\partial_\alpha\partial_\beta\right]^m\,.
\label{beginr}
\end{equation}
Here, $f_{m,n}^u$, $f_{m,n}^T$, $g^u_{m,n}$, $g^T_{m,n}$, $r^T_m$, $r^u_m$ are functions of temperature, but no derivative of temperature and $\hat P^{\mu\nu}$ is the projector perpendicular to $\hat{u}^\mu$, $\hat P^{\mu\nu} = \eta^{\mu\nu} + \hat u^\mu \hat u^\nu$.
Note at the level of  linearized analysis the two types of derivatives $(\hat{u}^\alpha\partial_\alpha)$ and $\left(\hat{P}^{\alpha\beta}\partial_\alpha\right)$ commute, so their relative ordering in the above functions does not matter.
 
In a rotational and translationally invariant equilibrium situation, the fluid profile reads
 \begin{equation}
 \bar u^\mu = \{1,0,0,0\}, \qquad \bar T = {\rm constant},
 \end{equation}
and the perturbed solution looks like
\begin{equation}
u_\mu^{(s)} = \bar u_\mu +\epsilon~ \delta u_\mu ~e^{-i\omega t + i\vec k\cdot\vec x},\qquad T^{(s)} = \bar T +\epsilon~ \delta T ~e^{-i\omega t + i\vec k\cdot\vec x}\,.
\end{equation}
Since both $u_\mu^{(s)}$ and $\bar u^\mu$ are unit normalized, it follows that $\delta u\cdot \bar u \sim {\cal O}(\epsilon^2)$ or $\delta u^\mu$ has the form
 $\delta u^\mu =\{0,\vec\beta\}$. We could express $\vec \beta$ as 
 \begin{equation}
 \vec\beta = \beta_k\left(\vec k\over k\right)+\vec \beta_\perp~~\text{where} ~~k = \sqrt{\vec k\cdot\vec k},~~\beta_k =\left(\vec\beta\cdot\vec k\over k\right),~~\text{and}~~ \vec k\cdot\vec\beta_\perp =0\,.
 \end{equation}
Substituting in equation \eqref{motstg} we find
\begin{equation}\label{deltapert}
\begin{split}
\Delta u_\mu &= \{0, \vec {\Delta u}\},\\
\text{where we define}~\vec {\Delta u}&=\Delta u_k \left(\vec k\over k\right) +\vec{ \Delta u_\perp} ~~\text{such that} ~~\vec{\Delta u}_\perp\cdot\vec k =0\,.\\
\end{split}
\end{equation}
Therefore, the shift functions are expressed as~\cite{Bhattacharyya:2024jxm}
\begin{gather}\label{flufra}
 \begin{bmatrix} \Delta T\\
 \Delta u_k\\
 \Delta u_\perp
 \end{bmatrix}
 = S_{ij} \,
    \begin{bmatrix}
    \delta T\\
    \beta_k\\
    \beta_\perp
    \end{bmatrix}\,,
   \end{gather}
   where
       \begin{footnotesize}
\begin{equation}
S_{ij}
 =\begin{bmatrix}
  -i\omega~ G_T[-i\omega,-k^2]-k^2R_T[-k^2]&ik ~G_u[-i\omega,-k^2]&0\\
  ik~ F_T[-i\omega,-k^2] &-i\omega ~F_u[-i\omega,-k^2]-k^2R_u[-k^2]&0\\
  0&0&-i\omega ~F_u[-i\omega,-k^2]-k^2R_u[-k^2]
   \end{bmatrix}\nonumber
   \end{equation}
       \end{footnotesize}
with $\Delta u_k$ and $\Delta u_\perp$ being defined in the same fashion as $\beta_k$ and $\beta_\perp$. 
The equation \eqref{flufra} determines the matrix $S_{ij},~~\{i,j\} =\{1,2,3\}$ (defined in equation \eqref{shiftlin}) for the special case of fluid dynamics. It is easy to compute the determinant of the matrix $\left[{\mathbf{1}} + S\right]$ which is expressed as
   \begin{equation}
   \begin{split}
   {\cal F}(-i\omega, k^2)&=Det[{\mathbf{1}} +S] \\
   &=(1-i\omega F_u-k^2R_u)\left[(1-i\omega G_T-k^2R_T)(1-i\omega F_u-k^2R_u) + k^2 F_T~G_u\right].
   \end{split}
   \label{fluidet}
   \end{equation}
   To avoid clutter, we have omitted the arguments of \( F_u \), \( F_T \), \( G_u \), \( G_T \), \( R_T \), and \( R_u \). Based on equation \eqref{finclude}, we conclude that under a general field redefinition, the dispersion polynomial acquires an additional factor, \( \mathcal{F} \). This factor is usually a polynomial in \( \omega \) and \( k^2 \), and as discussed in the previous section, it can introduce new modes into the system. These functions exhibit the following general properties:
   \begin{enumerate}
\item If the functions \( F_u \) and \( G_T \) are zero, and \( F_T \) and \( G_u \) are independent of the operator \( (u\cdot\partial) \) or \( (-i\omega) \) in Fourier space, then \( \mathcal{F} \) will not have any zeros for any value of \( \omega \). In such frame transformations, no new modes will be generated. However, depending on the specific forms of \( F_T \) and \( G_u \), it is possible that \( \mathcal{F} \) could have zeros for certain real values of \( k \).
\item If every term in \( F_u \) and \( G_T \) contains at least one factor of the operator \( (P^{\mu\nu}\partial_\mu\partial_\nu) \) or a factor of \( k^2 \) in Fourier space\footnote{Note that \( R_T \) and \( R_u \) are, by definition, independent of \( \omega \).}, then the frequency of these new modes will diverge as the momentum approaches zero \( (k \rightarrow 0) \).
\item If either \( F_u \) or \( G_T \) contains at least one term without a factor of \( (P^{\mu\nu}\partial_\mu\partial_\nu) \), then at least one new mode will emerge with a finite, non-zero frequency as \( k \rightarrow 0 \). This mode will resemble a genuine ``non-hydrodynamic'' mode.
\item None of the zeros of \( \mathcal{F} \) will take the form where \( \lim_{k \rightarrow 0} \omega(k) = 0 \). In other words, frame redefinitions will never generate new hydrodynamic modes. Thus, in Fourier space, if we focus solely on modes with small \( \omega \) and small \( k \), the redefinition of fluid variables will not affect the spectrum of hydrodynamic modes.
   \end{enumerate}
\section{Can we remove a mode from the spectrum via frame redefinition?}
   \label{sec:framerem}
We have thus far discussed how frame redefinition introduces new modes into the spectrum. For instance, in the previous section, we observed that the dispersion polynomial in `Frame-2' typically contains more non-hydrodynamic modes (arising from the zeros of the factor \( \mathcal{F} \) generated by the frame transformation) than in `Frame-1'.

Now, suppose we want to reverse this process—starting with the equations in `Frame-2' (using the hatted fluid variables \( \hat{u}^\mu \) and \( \hat{T} \)) and applying the inverse of the transformations from equations \eqref{fluidframe} and \eqref{motstg} to revert to the equations in `Frame-1' with fluid variables \( u^\mu \) and \( T \). This inverse transformation would naturally remove the extra factor \( \mathcal{F} \) from the dispersion polynomial of `Frame-2'. From the perspective of `Frame-2', this would effectively appear as a complete removal of a non-hydrodynamic mode from the spectrum through field redefinition.
   
The ability to completely absorb the information of a non-hydrodynamic mode through a field redefinition suggests that such a mode may be ``unphysical''. In other words, we expect that it should never be possible to fully erase the information of a truly physical non-hydrodynamic mode—one whose acausality or instability could invalidate the fluid model itself—through frame transformation. In this section, we will explore how to construct an ``inverse transformation'' that can remove a mode from the spectrum and examine the extent to which our expectation about the persistence of physical non-hydrodynamic modes holds true.

For convenience let us first restate how the dispersion polynomials in `Frame-1' (denoted as $P(\omega,k^2)$) and `Frame-2' (denoted as $\hat P(\omega, k^2)$) are related
\begin{eqnarray}
    \hat P(\omega, k^2) &=& P(\omega,k^2){{\cal F}(\omega,k^2)},\\
\text{where} \qquad {{\cal F}(\omega,k^2)}&\equiv&\underbrace{(1-i\omega F_u-k^2R_u)}_{\text{Shear Channel}}\underbrace{\left[(1-i\omega G_T-k^2R_T)(1-i\omega F_u-k^2R_u) + k^2 F_T~G_u\right]}_\text{Sound Channel}
\nonumber
\end{eqnarray}
and
\begin{eqnarray}
F_u &\equiv& \sum_{m,n} f_{m,n}^u \left[-i\omega\right]^m \left[-k^2\right]^n,~~~
F_T \equiv \sum_{m,n} f_{m,n}^T \left[-i\omega\right]^m \left[-k^2\right]^n\,,\nonumber\\
G_u &\equiv& \sum_{m,n} g^u_{m,n} \left[-i\omega\right]^m \left[-k^2\right]^n,~~G_T \equiv \sum_{m,n} g^T_{m,n} \left[-i\omega\right]^m \left[-k^2\right]^n\,,\nonumber\\
R_u &\equiv& \sum_{m} r^u_{m} \left[-k^2\right]^m\,,~\,~~\qquad \quad
R_T \equiv \sum_{m} r^T_{m} \left[-k^2\right]^m\,.
\label{rel12}
\end{eqnarray}
Given the isotropy of the background fluid profile and the constitutive relations, the linearized dynamics in the shear sector (where the velocity perturbation is perpendicular to the wave vector $\vec{k}$) and the sound sector (where the velocity perturbation is aligned with $\vec{k}$) will decouple. As a result, both $\hat{P}(\omega, k^2)$ and $P(\omega, k^2)$ must factorize cleanly into contributions from the shear and sound sectors. The same factorization applies to $\mathcal{F}(\omega, k^2)$.
In other words we could independently identify the factors as
\begin{eqnarray}
\text{Shear channel:}&~&\hat P_{\rm sh}(\omega, k^2) = P_{\rm sh}(\omega, k^2) (1-i\omega F_u-k^2R_u)\,,\label{shearsound}\\
\text{Sound channel:}&~&\hat P_{{\rm snd}}(\omega,k^2) = P_{{\rm snd}}(\omega, k^2)\left[(1-i\omega G_T-k^2R_T)(1-i\omega F_u-k^2R_u) + k^2 F_T~G_u\right],\nonumber
\end{eqnarray}
such that
\begin{equation}
    \hat P(\omega, k^2) = \hat P_{\rm sh}(\omega, k^2)\hat P_{{\rm snd}}(\omega,k^2),~~~P(\omega,k^2)=P_{\rm sh}(\omega, k^2) P_{{\rm snd}}(\omega, k^2).
\end{equation}
Further assume that $P(\omega,k^2)$  in shear channel has $N^{\rm sh}_1$ hydrodynamic modes with frequencies ~$\omega_{\rm sh}^a(k),~~a=\{1,2\cdots, N^{\rm sh}_1\}$, and $N^{\rm sh}_2$ non-hydrodynamic modes with frequencies ~${\mathfrak w}_{\rm sh}^a(k^2),~~a=\{1,2\cdots, N^{\rm sh}_2\}$ and similarly, in the sound channel it has $N^{{\rm snd}}_1$ hydrodynamic modes with frequencies ~$\omega_{{\rm snd}}^a(k),~~a=\{1,2\cdots, N^{{\rm snd}}_1\}$, and $N^{{\rm snd}}_2$ non-hydrodynamic modes with frequencies ~${\mathfrak w}_{{\rm snd}}^a(k^2),~~a=\{1,2\cdots, N^{{\rm snd}}_2\}$. Consequently, $P^{\rm sh}(\omega,k^2)$  and $ P^{{\rm snd}}(\omega, k^2)$ can be factorized as
\begin{equation}
\begin{split}
P^{\rm sh}(\omega,k^2) &=\left[\prod_{a=1}^{N^{\rm sh}_1}(\omega-\omega_{\rm sh}^{(a)})\right]\left[\prod_{a=1}^{N^{\rm sh}_2}(\omega-{\mathfrak w}_{\rm sh}^{(a)})\right],\\
P^{{\rm snd}}(\omega,k^2) &=\left[\prod_{a=1}^{N^{{\rm snd}}_1}(\omega-\omega_{{\rm snd}}^{(a)})\right]\left[\prod_{a=1}^{N^{{\rm {\rm snd}}}_2}(\omega-{\mathfrak w}_{{\rm {\rm snd}}}^{(a)})\right].
\end{split}
\end{equation}
Both $\omega^{(a)}_{{\rm sh}/{\rm {\rm snd}}}$ and ${\mathfrak w}_{{\rm sh}/{\rm {\rm snd}}}^{(a)}$ are typically complicated non-polynomial functions of $k$ with limits
$$\lim_{k\rightarrow0}\omega^{(a)}_{{\rm {\rm snd}}/{\rm sh}}(k)=0,~~~\lim_{k\rightarrow0}{\mathfrak w}^{(a)}_{{\rm sh}/{\rm {\rm snd}}}(k)=\text{some finite constant $c^{(a)}_{{\rm sh}/{\rm {\rm snd}}}$}~~~\forall a\,,$$
where we note that $\omega^{(a)}_{{\rm {\rm snd}}/{\rm sh}}(k)$ are hydrodynamic modes and ${\mathfrak w}^{(a)}_{{\rm sh}/{\rm {\rm snd}}}(k)$ are non-hydrodynamic modes.

Our goal is to find a frame transformation such that the transformed constitutive relation has no non-hydrodynamic modes and remains exactly equivalent to the original system in the hydrodynamic sector. This exact equivalence means that both the frequencies and the eigenvectors (or the profile of linearized perturbations satisfying the equations of motion) in the hydrodynamic sector should match perfectly before and after the frame transformation.

It is important to note that the allowed non-trivial profiles of the linearized perturbation also decompose cleanly into two subspaces: the two-dimensional space of the sound channel, spanned by the velocity perturbation in the direction of the wave-vector (\( \vec{k} \)) and the temperature perturbation, and the one-dimensional subspace of the shear channel, represented by the velocity perturbation in the direction perpendicular to \( \vec{k} \).

Since the hydrodynamic eigen-space in the shear channel is one-dimensional, once the frame transformation removes the shear non-hydrodynamic frequency from the spectrum without affecting the hydrodynamic one, the matching of the eigenvector is automatic. However, in the sound channel, the eigenvector is a special vector in the two-dimensional subspace of the temperature and longitudinal velocity perturbations, characterized by the ratio \( \mathcal{R} \) of these two components
\begin{equation}
    {\cal R}(\omega,k) = \frac{\delta T}{\delta u_k}\,.
    \label{eq:R_ratio}
\end{equation}
In other words, in the sound channel, after removing the non-hydrodynamic mode, we must also ensure that not only the hydrodynamic frequencies but also \( \mathcal{R} \) remain unchanged before and after the frame transformation.

For convenience let us first quote the expression of ${{\cal F}(\omega,k^2)}$, the extra factor in the dispersion polynomial generated due to the frame transformation
\begin{equation}\label{eq:quote}
\begin{split}
&{{\cal F}(\omega,k^2)}\equiv\underbrace{(1-i\omega F_u-k^2R_u)}_{\text{Shear Channel}}\underbrace{\left[(1-i\omega G_T-k^2R_T)(1-i\omega F_u-k^2R_u) + k^2 F_T~G_u\right]}_\text{Sound Channel}\,.
\end{split}
\end{equation}
Now the above discussion about removal of non-hydrodynamic modes lead to the following equations on the frame transformation
\begin{equation}\label{eq:frametrans}
\begin{split}
&{(1-i\omega F_u-k^2R_u)}\bigg[\prod_{a=1}^{N^{\rm sh}_2}(\omega-{\mathfrak w}_{\rm sh}^a)\bigg]=C_{\rm sh}\,,\\
&\left[(1-i\omega G_T-k^2R_T)(1-i\omega F_u-k^2R_u) + k^2 F_T~G_u\right]\bigg[\prod_{a=1}^{N^{{\rm {\rm snd}}}_2}(\omega-{\mathfrak w}_{{\rm {\rm snd}}}^a)\bigg]=C_{{\rm {\rm snd}}}\,,\\
\text{where}~~&C_{\rm sh} = \lim_{k\rightarrow 0}\left(\prod_{a=1}^{N_2^{\rm sh}}\left[-{\mathfrak w}^a_{\rm sh}\right]\right),~~~~~C_{{\rm {\rm snd}}} = \lim_{k\rightarrow 0}\left(\prod_{a=1}^{N_2^{{\rm {\rm snd}}}}\left[-{\mathfrak w}^a_{{\rm {\rm snd}}}\right]\right)\,.
\end{split}
\end{equation}
Clearly, if the functions $F_u,~ F_T,~G_u,~G_T,~R_T,~R_u$ are finite polynomials in $\omega$ and $k^2$, the condition  \eqref{eq:frametrans} could never be satisfied. But if we allow derivative expansion  (in Fourier space, an expansion around $\omega=0, ~k=0$) up to infinite order, then we could, in principle remove,  all non-hydrodynamic modes from the dispersion polynomial by an appropriate choice of the expansion coefficients $f_{m,n}^u ,~~f_{m,n}^T,~~g^u_{m,n},~~g^T_{m,n},~~r^T_m,~~r^u_m$. It also follows that the radius of convergence of this infinite order frame transformation in the Fourier space will be determined by the location of the lowest (nearest to the origin in the multi-dimensional space of real momenta and complex frequencies) non-hydrodynamic mode.

Finally, suppose we normalize the perturbed solution in the sound channel in the following way
$$u_\mu^{(s)} = \bar u_\mu +\epsilon~ \{0,{\vec k/ k}\} ~e^{-i\omega t + i\vec k\cdot\vec x},~~T^{(s)} = \bar T +\epsilon~ \delta T ~e^{-i\omega t + i\vec k\cdot\vec x},$$
where  $\delta T$ is given by the expression ${\cal R}(\omega,k)$ before the frame transformation and $\hat {\cal R}(\omega,k)$ after the frame transformation.
Then the equivalence of the hydrodynamic sector demands that
\begin{equation}\label{eq:frameeigen}
\begin{split}
{\cal R}(\omega^a_{{\rm {\rm snd}}}, k) = \hat{\cal R}(\omega^a_{{\rm {\rm snd}}},k)~~~~\forall a = \{1,2,N_1^{{\rm {\rm snd}}}\}\,.
\end{split}
\end{equation}
Note that the detailed structure of ${\hat{\cal R}}(\omega^a_{{\rm {\rm snd}}},k)$ will depend on the functions involved in frame transformations namely $\{F_T,G_T,G_u,F_u,R_T,R_u\}$.

Now let us discuss whether we can consistently solve these set of coupled  non-linear equations (\eqref{eq:frametrans} and \eqref{eq:frameeigen}) and to what extent the solution would be unique.
\begin{itemize}
    \item Naively, it seems that we have $(N_1^{{\rm {\rm snd}}} +2)$  number of equations  for 6 unknown functions. However, note that $R_T$ and $R_u$ could only be functions of $k$ and do not depend on $\omega$.
    \item So if we take $\omega\rightarrow0$ limit in the two equations of \eqref{eq:frametrans}, it effectively gives two more equations involving $R_T$, $R_u$, and $[F_T ~G_u]_{(0)}$ (which is independent of $\omega$ and denoted as $[F_T ~G_u]_{(0)}\equiv\lim_{\omega\rightarrow0}\left[F_T(\omega,k)G_u(\omega,k)\right]$).
    \item In fact using these two new equations we could solve for $R_T$ and $R_u$ uniquely in terms of $[F_T ~G_u]_{(0)}$ and other known functions of $k$ like ${\mathfrak w}^a_{{\rm {\rm snd}}}(k)$ and ${\mathfrak w}^a_{\rm sh}(k)$.
    \item In the next step, we substitute the solution of $R_T$ and $R_u$ and switch on $\omega$ in both the equations of \eqref{eq:frametrans}. Now we can solve for $F_u$ and $G_T$ uniquely in terms of the product $F_T~G_u$ and $[F_T~G_u]_{(0)}$.
\end{itemize}
 At this stage both the equations in \eqref{eq:frametrans} will be solved for all values of $\omega$ and $k$.
Next we have $N_1^{{\rm {\rm snd}}}$ equations from \eqref{eq:frameeigen} to solve for $F_T$ and $G_u$. For a typical uncharged hydrodynamic theory like the ones that we are considering here $N_1^{{\rm snd}}=2$ and so we have precisely 2 equations for the 2 remaining unknown functions $F_T$ and $G_u$. However, the equations in \eqref{eq:frameeigen} are all evaluated at $\omega = \omega^a_{{\rm snd}} (k)$. Therefore, solving these equations,  we can never uniquely determine $F_T$ and $G_u$ for all values of $\omega$, e.g., we could always add two separate arbitrary functions of the product $\prod_a\left(\omega - \omega^a_{{\rm snd}}\right)$ to any solutions for $F_T$ and $G_u$ and they will continue to solve the equations \eqref{eq:frameeigen}.

So to summarize,  barring the small ambiguity in $F_T$ and $G_u$ described above, we could almost uniquely solve for the functions appearing in the linearized frame transformation by demanding that the spectrum in the transformed frame contains only the hydrodynamic modes\footnote{Note that, had we not introduced the functions $R_T(k)$ and $R_u(k)$ in the original structure of the linearized frame transformation i.e., equation \eqref{motstg}, $F_u$ and $G_T$ must diverge in the limit $\omega\rightarrow0$ in order to satisfy the set of equations \eqref{eq:frametrans} and \eqref{eq:frameeigen}}.
\section{Field redefinition and the non hydrodynamic mode in BDNK theory}\label{sec:BDNK}
In this section we shall apply the algorithm described before to remove the non-hydrodynamic modes in the BDNK theory~\cite{Bemfica:2017wps,Kovtun:2019hdm}. 
For a conformal system (energy density $\sim T^4$), the stress tensor of the theory is given as follows\footnote{Here we have assumed the stress tensor described in equation \eqref{bdnkstress} is an exact stress tensor and does not have any higher order derivative corrections. The purpose of this section is to show how we could remove the non-hydrodynamic modes. The BDNK model has been used to illustrate the procedure and we have chosen to be agnostic about the physical validity of a first order theory.}
\begin{equation}
T^{\mu\nu} = T^4\bigg\{(4 u^\mu u^\nu + \eta^{\mu\nu})
 +   4  \left[ \theta \left(u^\mu q^\nu + u^\nu q^\mu\right) +    \chi \,E~(4 u^\mu u^\nu + \eta^{\mu\nu})- 2 \lambda~ \sigma^{\mu\nu} \right]\bigg\}\,,
 \label{bdnkstress}
\end{equation}
where
\begin{equation}
\begin{split}
q^\mu&\equiv \left[(u\cdot\partial)u^\mu +  \left(P^{\mu\nu} \partial_\nu T\over T\right)\right],\\
E &\equiv \left[{(u\cdot \partial) T\over T} + \frac{\left(\partial\cdot u\right)}{3} \right],\\
\sigma^{\mu\nu} &\equiv \left({P^{\mu\alpha} P^{\nu\beta} +P^{\mu\beta} P^{\nu\alpha} \over 2 }-{P^{\mu\nu}P^{\alpha\beta}\over 3} \right)\left(\partial_\alpha u_\beta\right)\,,
\end{split}
\end{equation}
and $\{\chi,\theta,\lambda\}$ are arbitrary constants 
which, for a system with no conserved charges, are functions of $T$ only.
The dispersion polynomial resulting from the conservation of stress tensor is as follows\footnote{Here we have re-scaled the transport coefficients ($\theta$, $\chi$ and $\lambda$) with appropriate factors of $T$ so that the combinations (transport coefficients $\times$ frequency($\omega$)) or (transport coefficients $\times$ momentum ($k$)) are dimensionless.}
\begin{equation}\label{eq:disperseBDNK}
\begin{split}
P_\text{BDNK}(\omega,k) =&
 \underbrace{\left[\omega  (1- i\theta~ \omega )+i\lambda k^2  \right]}_{{\cal P}_{shear}}\\
 &\times \underbrace{\bigg\{\left[3 \omega 
   (1-i\theta~ \omega)-i\chi~ k^2\right] \left[3\omega  (1-i\chi~ \omega )-i(\theta-4\lambda)~ k^2\right]+3i k^2 [1-i(\chi+\theta)~\omega ]^2 \bigg\}}_{{\cal P}_{sound}}.
\end{split}
\end{equation}
This theory has, in the shear channel, one hydrodynamic and one non-hydrodynamic mode, whereas, in the sound channel, it has two hydrodynamic and two non-hydrodynamic modes. Since we would like to determine the frame transformation in an expansion in $\omega$ and $k$, for our purpose it is enough to know the modes in a similar expansion in $k$ (up to ${\cal O}(k^4)$)
\begin{equation}\label{modesBDNK}
\begin{split}
{\mathfrak w}_{\rm sh}=&-{i\over\theta} + i\lambda~ k^2 \,,\\
{\mathfrak w}^{(1)}_{{\rm snd}}=&-{i\over\theta} + {i\over 3}\left[4\lambda -{\theta\chi\over(\theta-\chi)}\right]k^2 \,,\\
{\mathfrak w}^{(2)}_{{\rm snd}}=&-{i\over\chi} + {i\over 3}\left(\theta\chi\over\theta-\chi\right)k^2 \,,\\
\omega_{\rm sh}=&-i\lambda~k^2 \,,\\
\omega^{(1)}_{{\rm snd}}=&{k\over\sqrt{3}}- \left(2i\lambda\over 3\right) k^2 - \left(2\lambda^2\over 3\sqrt{3}\right) k^3 \,,\\
\omega^{(2)}_{{\rm snd}}=&-{k\over\sqrt{3}}- \left(2i\lambda\over 3\right) k^2 + \left(2\lambda^2\over 3\sqrt{3}\right) k^3 \,,\\
 {\cal R}(\omega,k) =&\,  k\left[i + (\theta+\chi)\omega\over k^2\theta +3\omega (i +\chi\omega)\right].
 \end{split}
\end{equation}
Once we have the expressions for the modes, we substitute them in equation \eqref{eq:frametrans} and \eqref{eq:frameeigen}  and proceed to solve. Our strategy is as follows:
\begin{itemize}
\item We first take $\omega\rightarrow 0 $ limit  in the first equation of \eqref{eq:frametrans} and then solve for $R_u(k)$. Formally, the solution takes the form
\begin{equation}\label{sol:ru}
\begin{split}
R_u(k)=\left(1\over k^2\right) \left( {\mathfrak w}_{\rm sh} +{i\over\theta}\over {\mathfrak w}_{\rm sh}\right).
\end{split}
\end{equation}
\item We then substitute the solution for $R_u(k)$ in the same equation and turn on $\omega$ to solve for $F_u(\omega,k)$
\begin{equation}\label{sol:fu}
\begin{split}
F_u(\omega,k)=-{1/\theta\over {\mathfrak w}_{\rm sh}(\omega-{\mathfrak w}_{\rm sh})}.
\end{split}
\end{equation}
\item Next we come to the second equation of \eqref{eq:frametrans} and substitute the solutions for $F_u$ and $R_u$.
\item As before we first take the $\omega\rightarrow 0$ limit of this  equation and solve for $R_T(k)$
\begin{equation}\label{sol:rt}
\begin{split}
R_T(k)&=i\theta~{\mathfrak w}_{\rm sh}\left[F_T~G_u\right]_{(0)}+\left(1\over k^2\right) \left[1+{i\over\chi} \left({\mathfrak w}_{\rm sh}\over {\mathfrak w}_{{\rm snd}}^{(1)}{\mathfrak w}_{{\rm snd}}^{(2)}\right)\right],\\
\text{where}~&\left[F_T~G_u\right]_{(0)}\equiv\lim_{\omega\rightarrow 0}\left[ F_T(\omega,k)G_u(\omega,k)\right].
\end{split}
\end{equation}
\item We substitute the above solution in the same equation, turn on $\omega$ and solve for $G_T(\omega,k)$
\begin{equation}\label{sol:gt}
\begin{split}
G_T(\omega,k)&=\theta~{\mathfrak w}_{\rm sh}\left(k^2\over\omega\right)\left(F_T G_u -\left[F_T~G_u\right]_{(0)}\right)-\theta ~k^2(F_T ~G_u)\\
&~~-{1\over\chi} \left[{\mathfrak w}_{{\rm snd}}^{(1)}{\mathfrak w}_{{\rm snd}}^{(2)}+{\mathfrak w}_{\rm sh}\left(\omega -{\mathfrak w}_{{\rm snd}}^{(1)}-{\mathfrak w}_{{\rm snd}}^{(2)}\right)\over {\mathfrak w}_{{\rm snd}}^{(1)}{\mathfrak w}_{{\rm snd}}^{(2)}\left(\omega -{\mathfrak w}_{{\rm snd}}^{(1)}\right)\left(\omega -{\mathfrak w}_{{\rm snd}}^{(2)}\right)\right].
\end{split}
\end{equation}
\item At this stage we have precisely two unknown functions $F_T(\omega, k)$ and $G_u(\omega,k)$ and two equations from \eqref{eq:frameeigen} evaluated at the  frequencies of the two hydrodynamic sound modes $\omega_{{\rm snd}}^{(1)}$ and $\omega_{{\rm snd}}^{(2)}$. The precise structure (after we have determined the expression $\hat{\cal R}(\omega,k)$ - the ratio of the longitudinal velocity and the temperature perturbation in the transformed frame) is as follows
\begin{equation}
\begin{split}
\hat{\cal R}(\omega,k) & ={ k\left[G_u\left(\theta k^2  +3\chi\omega^2 + 3 i\omega \right) +i\left(1 - i\omega F_u-k^2R_u\right)\left(i + \theta\omega+\chi\omega\right)\right]\over k^2\left(i + \theta\omega+\chi\omega\right)F_T + i\left(1-k^2R_T -i\omega G_T\right)\left(\theta k^2  +3\chi\omega^2 + 3 i\omega \right)},\\
{\cal R}(\omega,k) &=  k\left[i + (\theta+\chi)\omega\over k^2\theta +3\omega (i +\chi\omega)\right],\\
{\cal R}(\omega_{{\rm snd}}^{(1)},k) &=\hat{\cal R}(\omega_{{\rm snd}}^{(1)},k),\\
{\cal R}(\omega_{{\rm snd}}^{(2)},k) &=\hat{\cal R}(\omega_{{\rm snd}}^{(2)},k).
\label{pres}
\end{split}
\end{equation}
There are couple of points to be noted here:
\begin{itemize}
\item Both the equations in \eqref{pres} have to be satisfied only at the frequencies of the hydrodynamic sound modes. All explicit dependence on $\omega$ has to be replaced by $\omega^{(1)}_{{\rm snd}}(k)$ and $\omega^{(2)}_{{\rm snd}}(k)$. However, by solving these equations we could never determine the unknown functions $F_T$ and $G_u$ at all values of $\omega$ and $k$.
\item But, our goal is to show that there exists at least one frame transformation that  removes the non-hydrodynamic modes. If we simply assume that both $F_T$ and $G_u$ are independent of $\omega$, equation \eqref{pres} will allow us to uniquely solve these two functions, thus generating one frame transformation suitable for our purpose. After imposing this assumption, the solution for $R_T$ and $G_T$ simplifies
\begin{equation}\label{rtgtsimp}
\begin{split}
R_T(k)&=i\theta~{\mathfrak w}_{\rm sh}F_T~G_u+\left(1\over k^2\right) \left[1+{i\over\chi} \left({\mathfrak w}_{\rm sh}\over {\mathfrak w}_{{\rm snd}}^{(1)}{\mathfrak w}_{{\rm snd}}^{(2)}\right)\right]\,,\\
G_T(\omega,k)&=-\theta ~k^2(F_T ~G_u)-{1\over\chi} \left[{\mathfrak w}_{{\rm snd}}^{(1)}{\mathfrak w}_{{\rm snd}}^{(2)}+{\mathfrak w}_{\rm sh}\left(\omega -{\mathfrak w}_{{\rm snd}}^{(1)}-{\mathfrak w}_{{\rm snd}}^{(2)}\right)\over {\mathfrak w}_{{\rm snd}}^{(1)}{\mathfrak w}_{{\rm snd}}^{(2)}\left(\omega -{\mathfrak w}_{{\rm snd}}^{(1)}\right)\left(\omega -{\mathfrak w}_{{\rm snd}}^{(2)}\right)\right]\,.
\end{split}
\end{equation}
\end{itemize}
\item Finally even after assuming that $F_T$ and $G_u$ to be independent of $\omega$, the equation \eqref{pres} is a set of two  coupled non-linear algebraic equations (non-linearity arises due to the fact that $R_T$ depends on the product of $F_T$ and $G_u$). Though straightforward to solve, the solution to equation \eqref{pres} turns out to be a bit cumbersome if we want to write the exact expressions for $F_T$ and $G_u$ as we did for the other functions $F_u$, $R_u$, $R_T$ and $G_T$. Below we are briefly describing the steps we used to solve them.
\begin{enumerate}
\item Using \eqref{pres} we first solve for $F_T$ and $G_u$ in terms of $F_u$, $R_u$, $R_T$ and $G_T$, ignoring the fact that $R_T$ and $G_T$ themselves depend on the product of  $F_T$ and $G_u$. We get the following solution
\begin{equation}\label{newsolFT}
\begin{split}
F_T(k)=\frac{1}{k^2}\left(A^{(1)}B^{(1)}A^{(2)}B^{(2)}\over \left[A^{(2)}B^{(1)}\right]^2 -\left[A^{(1)}B^{(2)}\right]^2\right)\Bigg\{&\left(A^{(2)}\over B^{(2)}\right)\left[ik^2(R_T -R_u)+\omega_{{\rm snd}}^{(1)}\left(F_u^{(1)} -G_T^{(1)}\right)\right]\\
-\left(A^{(1)}\over B^{(1)}\right)&\left[ik^2(R_T -R_u)+\omega_{{\rm snd}}^{(2)}\left(F_u^{(2)} -G_T^{(2)}\right)\right]\Bigg\},
\end{split}
\end{equation}
\begin{equation}\label{newsolGu}
\begin{split}
G_u(k)=\left(A^{(1)}B^{(1)}A^{(2)}B^{(2)}\over \left[A^{(2)}B^{(1)}\right]^2 -\left[A^{(1)}B^{(2)}\right]^2\right)\Bigg\{&\left(B^{(2)}\over A^{(2)}\right)\left[ik^2(R_T -R_u)+\omega_{{\rm snd}}^{(1)}\left(F_u^{(1)} -G_T^{(1)}\right)\right]\\
-\left(B^{(1)}\over A^{(1)}\right)&\left[ik^2(R_T -R_u)+\omega_{{\rm snd}}^{(2)}\left(F_u^{(2)} -G_T^{(2)}\right)\right]\Bigg\},
\end{split}
\end{equation}
where
\begin{equation}\label{nott}
\begin{split}
&A^{(1,2)} \equiv  k^2\theta +3~\omega_{{\rm snd}}^{(1,2)} ~\left[i +\chi~\omega_{{\rm snd}}^{(1,2)}\right], \quad B^{(1,2)} \equiv i + (\theta+\chi)~\omega_{{\rm snd}}^{(1,2)},\\
&F_u^{(1,2)}\equiv F_u(\omega_{{\rm snd}}^{(1,2)},k),\quad G_T^{(1,2)}\equiv G_T(\omega_{{\rm snd}}^{(1,2)},k).
\end{split}
\end{equation}
\item However, equations \eqref{newsolFT} and \eqref{newsolGu} are not full solutions for $F_T$ and $G_u$ since both the functions $R_T$ and $G_T$ contain the product of $F_T$ and $G_u$ (see equations \eqref{sol:gt} and \eqref{sol:rt}).
\item We take the product of equations \eqref{newsolFT} and \eqref{newsolGu} and using equations \eqref{sol:gt} and \eqref{sol:rt} we finally get a quadratic equation for the product $(F_T~G_u)$.
\item Solving this equation we first get a solution for the product $(F_T~G_u)$ as function of $k$ and substituting it back in equations \eqref{sol:rt} ans \eqref{sol:gt} we get explicit solutions for $R_T$ and $G_T$.
\item Once we know the explicit expressions for $R_T$ and $G_T$,  \eqref{newsolFT} and \eqref{newsolGu} give the full solution for $F_T$ and $G_u$.
\end{enumerate}
\item Though straightforward to carry out in computers, it turns out that the intermediate expressions are too cumbersome to present here.
The final leading solutions are as follows~\cite{Bhattacharyya:2024jxm}
\begin{equation}\label{ftgu}
\begin{split}
G_u(k)& =g_0 +  {\cal O}(k^2)\,,\qquad
  F_T(k)=3g_0 + \chi-\theta +  {\cal O}(k^2)\,,\\
\text{where}~g_0&\equiv \frac{1}{6} \left(-\sqrt{5 \theta ^2-16 \theta  \lambda -6 \theta  \chi +16 \lambda ^2+8 \lambda  \chi -3 \chi ^2}+\theta +4
   \lambda -\chi \right)\,.
\end{split}
\end{equation}
\item Once we have solved for all the functions appearing in the frame transformation in the Fourier space, we can use the following simple replacement to convert them to expressions in position space
\begin{equation}\label{replace}
\begin{split}
\text{Replacement}: ~~\omega\rightarrow-i (u\cdot\partial),~~~k^2\rightarrow -P^{\mu\nu}\partial_\mu\partial_\nu\,.
\end{split}
\end{equation}
\end{itemize}
At this stage, it is crucial to emphasize that although we have presented a formal exact solution for \( F_u \), \( R_u \), \( R_T \), and \( G_T \), these solutions should always be understood as expansions in non-negative powers of \( \omega \) and \( k \). Polynomials of \( k \) and \( \omega \) in the denominator simply imply an infinite power series with a radius of convergence determined by the zeros of those polynomials.

Our entire analysis is based on the derivative expansion, and the above replacement of \(\{\omega, k^2\}\) in terms of derivatives does not make sense if we allow powers of \( \omega \) and \( k \) in the denominator. It is important to note that all the functions \( F_u \), \( F_T \), \( G_u \), \( G_T \), \( R_T \), and \( R_u \) are finite when taking the limit \(\omega \rightarrow 0\) and/or \( k \rightarrow 0\) in any order. This ensures that after the expansion and replacement, there will not be any negative power of \((u \cdot \partial)\) or \((P_\mu^\nu \partial_\nu)\).
For instance, while $R_u(k)$ or $R_T(k)$ might appear to diverge at $k \to 0$ due to the $1/k^2$ factors, they actually don't because when the infinite series of the non-hydrodynamic modes are substituted in, the terms in the numerator cancel out.

Similarly, it is essential that the final solution for the frame transformation depends only on even powers of \( k \) so that we can apply the replacement without using any non-analytic expressions like the square root of a derivative. As mentioned before, since we are only concerned with the spectrum of linearized perturbations, the order of the two derivatives \((u \cdot \partial)\) and \( P^\mu_\nu \partial_\mu \) is irrelevant.

Since the frame transformation includes terms of all orders in the derivative expansion, the resultant stress tensor in the transformed frame also contains terms of all orders in the derivative expansion.
\section{`Artificial' non-hydro modes under inverse field redefinition}
\label{sec:artificial}
In the previous section, we demonstrated that non-hydrodynamic modes could in principle be removed by an all-order inverse field redefinition. However, such an all-order field redefinition is necessary regardless of whether the non-hydrodynamic mode in question is a physical one or an artificial mode generated solely due to frame transformation. One way to understand this is by examining the equations of frame transformation, \(\eqref{fluidframe}\). These equations describe how fluid variables in `Frame-1' (\(u^\mu\) and \(T\)) can be expressed in terms of the fluid variables (\(\hat{u}^\mu\) and \(\hat{T}\)) in `Frame-2'. The dispersion relation in `Frame-2' includes non-hydrodynamic modes that are artefacts of this frame transformation. Removing these artificial modes from the dispersion relation is equivalent to reverting to a description in `Frame-1', or an inversion of the relation in \(\eqref{fluidframe}\)—requiring us to express \(\hat{u}^\mu\) and \(\hat{T}\) in terms of \(u^\mu\) and \(T\). Even if equation \(\eqref{motstg}\) and the RHS of \(\eqref{fluidframe}\) contain a finite number of derivatives, their inversion cannot be expressed solely with terms containing a finite number of derivatives acting on the fluid variables in `Frame-1'.

Thus, the form of this inverse field redefinition does not suffice to distinguish between these two types of non-hydrodynamic modes. To further investigate this issue, we will apply the algorithm of inverse frame description to a case where we know that the non-hydrodynamic mode is indeed an artificial one.

Consider the following stress tensor for an uncharged fluid\footnote{We would like to emphasize that the stress tensor described in equation \eqref{pseudo} is just a toy example to illustrate the issue of `artificial' non-hydro mode. There is no claim that such a stress tensor corresponds to any physical fluid or its conservation imposes any well-posed initial value problems for the fluid variables.}
\begin{equation}\label{pseudo}
\begin{split}
T^{\mu\nu} &= T^4\bigg\{(4 u^\mu u^\nu + \eta^{\mu\nu}) + 4\left[c\, \left(u^\mu a^\nu + u^\nu a^\mu\right) - 2 \lambda \left(\sigma^{\mu\nu}  + c~\tilde\sigma^{\mu\nu}\right)\right]\bigg\}\,,
\end{split}
\end{equation}
where
\begin{equation}
\begin{split}
a^\mu&\equiv (u\cdot\partial)u^\mu\,,~~~~~\sigma^{\mu\nu} \equiv \left({P^{\mu\alpha} P^{\nu\beta} +P^{\mu\beta} P^{\nu\alpha} \over 2 }-{P^{\mu\nu}P^{\alpha\beta}\over 3} \right)\left(\partial_\alpha u_\beta\right)\,,\\
\tilde \sigma^{\mu\nu} &\equiv\left({P^{\mu\alpha} P^{\nu\beta} +P^{\mu\beta} P^{\nu\alpha} \over 2 }-{P^{\mu\nu}P^{\alpha\beta}\over 3} \right)\left(\partial_\alpha a_\beta\right)\,,\\
\end{split}
\end{equation}
and $\{c,\lambda\}$ are arbitrary constants.

The dispersion polynomial for a fluid with such a stress tensor is given by (throughout this section, we have chosen units so that the equilibrium temperature $T_0$ is set to one)
\begin{equation}\label{distoy}
\begin{split}
P(\omega,k) &= \underbrace{\left(\omega +i\lambda~ k^2\right)}_{{\cal F}_1}\underbrace{\left( 3\omega^2- k^2 + 4 i \lambda~\omega k^2\right) }_{{\cal F}_2}\underbrace{( 1- i c~\omega)^2}_{{\cal F}_3}\,.
\end{split}
\end{equation}
The factors ${\cal F}_1$ and ${\cal F}_2$ contain the usual sound and shear hydrodynamic modes of the theory and it is the factor ${\cal F}_3$ that corresponds to the non-hydrodynamic mode.
We would like to remove ${\cal F}_3$ from the dispersion relation via inverse field redefinition. According to the discussion in section (\ref{sec:framerem}), in the transformed frame, the dispersion relation has to have the form
\begin{equation}\label{distoytrans}
\begin{split}
\hat P(\omega,k) &= \underbrace{\left(\omega +i\lambda~ k^2\right)}_{{\cal F}_1}\underbrace{\left( 3\omega^2- k^2 + 4 i \lambda~\omega k^2\right) }_{{\cal F}_2}\underbrace{( 1- i c~\omega)^2}_{{\cal F}_3}{\cal F}(\omega,k^2)\,,\\
\text{where}&~~ {\cal F} (\omega,k^2) =( 1- i c~\omega)^{-2} = \sum_{n=0}^\infty  (n+1) (ic~\omega)^n \,.
\end{split}
\end{equation}
Clearly, in the complex $\omega$ plane the series expansion of ${\cal F}$ has a radius of convergence as $\omega_{\rm boundary} = 1/|c|$. In this particular example, it is easy to find one  frame transformation that will lead to the above expansion of ${\cal F}$
\begin{equation}\label{simpfram}
\hat{u}^\mu =  u^\mu + \Delta \hat{u}^\mu (u), \qquad
\Delta \hat{u}^\mu = \sum_{n=1}^\infty (-1)^n c^n  (u\cdot\partial)^n  u^\mu
\end{equation}
Note again the expansion in $\Delta \hat{u}^\mu$ has the same radius of convergence as that of ${\cal F}$.

Now we shall substitute this frame redefinition in the stress tensor \eqref{pseudo}. The various expression that appears in the stress tensor would transform in the following way
\begin{equation}\label{transsimp}
\begin{split}
u^\mu u^\nu &= \hat u^\mu \hat u^\nu + \bigg[\hat u^\mu\sum_{n=1}^\infty (-1)^n c^n  (\hat u\cdot\partial)^n \hat u^\nu +\{\mu\leftrightarrow\nu\}\bigg],\\
\left(c ~u^\mu a^\nu +\{\mu\leftrightarrow\nu\}\right)&= \bigg[c~\hat u^\mu(\hat u\cdot\partial)\hat u^\nu + \hat u^\mu\sum_{n=1}^\infty (-1)^{n} c^{n+1}  (\hat u\cdot\partial)^{n+1} \hat u^\nu +\{\mu\leftrightarrow\nu\}\bigg],\\
&= -\bigg[ \hat u^\mu\sum_{n=1}^\infty (-1)^{n} c^{n}  (\hat u\cdot\partial)^{n} \hat u^\nu +\{\mu\leftrightarrow\nu\}\bigg].
\end{split}
\end{equation}
Note that under a frame transformation, the additional terms generated from the \(u^\mu u^\nu\) term in the stress tensor precisely cancel the additional terms generated from \(\left( c~u^\mu a^\nu\right)\). A similar pattern is observed in the transformation of the \(\sigma^{\mu\nu}\) and \(\tilde{\sigma}^{\mu\nu}\) terms.

In summary, to eliminate a non-hydrodynamic mode, the fields must be transformed to all orders in the derivative expansion. However, if the mode in question is an `artificial' one, most of the higher-order terms in the resultant stress tensor will cancel out (as shown in the calculations in this section). Consequently, even after removing the non-hydrodynamic mode, the stress tensor will retain a finite number of terms and remain convergent in Fourier space. In other words, the information of such artificial non-hydrodynamic modes can be completely eliminated from the stress tensor through appropriate field redefinition.

Conversely, if the mode is physical, an infinite-order field redefinition will generate a stress tensor with an infinite number of terms (see section \ref{sec:BDNK}). In Fourier space, the information about the physical non-hydrodynamic mode will be embedded in the radius of convergence of this infinite series.
\section{A `special second order correction' to first order BDNK theory }
\label{sec:special_correction}
Let us call any frame transformation performed to remove a particular non-hydrodynamic mode an `Inverse Frame Transformation.' A non-hydrodynamic mode is considered artificial or an artefact of frame transformation if we can find at least one inverse frame transformation that can eliminate the information of that particular mode by generating a stress tensor with a finite number of terms. In such a case, despite applying an infinite-order inverse frame transformation, the resultant stress tensor will have a finite number of terms due to some cancellations, as seen in the previous section.

In section \ref{sec:setup}, we observed that any non-hydrodynamic mode(s) generated by a frame transformation must originate from a factor matrix of the form \((\mathbf{1} + S)\) multiplying the linearized equations of motion, with a specific structure for \(S\) as described in equation \eqref{flufra}. This imposes several constraints on the structure of an `artefact' type non-hydrodynamic mode. One major constraint is the structure of the dispersion polynomial. Any non-hydrodynamic mode generated purely by frame transformation will always come from a separate factor of the form \(\mathcal{F}\) as given in equation \eqref{eq:quote}. When this is not the case, as we have seen in the BDNK theory (see equation \eqref{eq:disperseBDNK}), we can safely conclude that the non-hydrodynamic modes are physical (i.e., not artefacts of frame transformation).

In section \ref{sec:BDNK}, we removed the non-hydrodynamic modes of the BDNK theory by applying a specific inverse transformation. This was done in such a way that the hydrodynamic sector of the resultant infinite-order theory remained exactly equivalent to that of the original BDNK theory. From the structure of the frame transformation, we can infer that the information about the non-hydrodynamic modes is still hidden in the radius of convergence of the `transformed' infinite-order stress tensor.

In this section, we will consider another `toy example' of a stress tensor that is nearly identical to the BDNK stress tensor except for a very specific second-order correction. We will demonstrate that this `specially' corrected BDNK theory is equivalent to a frame-transformed first-order relativistic Navier-Stokes stress tensor, which is known to have many pathologies. In other words, the non-hydrodynamic modes in the BDNK theory are not artefacts of frame transformation, provided there is no second-order correction to the BDNK stress tensor of the specific structure described here (see equation \eqref{pseudo2}). If we view the BDNK stress tensor as an approximate one, correct only up to the first order in the derivative expansion, then it is always possible to remove all the novelties of the BDNK stress tensor by simply assuming the existence of some appropriate second-order terms.

Consider the following  stress tensor\footnote{We would like to emphasize that this is just a toy example to illustrate the diagnosis of artificial modes. This stress tensor does not represent any real physical system.}
\begin{equation}\label{pseudo2}
T^{\mu\nu} = 
T^4\bigg\{(4 u^\mu u^\nu + \eta^{\mu\nu})
 +  4 \left[\left(u^\mu q^\nu + u^\nu q^\mu\right) +  E~(4 u^\mu u^\nu + \eta^{\mu\nu})- 2 \lambda \, \sigma^{\mu\nu}- 2 \lambda \,\tilde\sigma^{\mu\nu}\right]\bigg\}\,,
\end{equation}
\begin{equation}
\begin{split}
\text{where} \quad
q^\mu&\equiv a_1 (u\cdot\partial)u^\mu +  a_2 \frac{\left(P^{\mu\nu} \partial_\nu T\right)}{T},\\
E &\equiv \left[ b_1 \frac{(u\cdot \partial)T}{T}  + b_2 \frac{1}{3} \left(\partial \cdot u \right)\right]\,, \\
\sigma^{\mu\nu} &\equiv \left({P^{\mu\alpha} P^{\nu\beta} +P^{\mu\beta} P^{\nu\alpha} \over 2 }-{P^{\mu\nu}P^{\alpha\beta}\over 3} \right)\left(\partial_\alpha u_\beta\right)\,,\\
\tilde \sigma^{\mu\nu} &\equiv  \left({P^{\mu\alpha} P^{\nu\beta} +P^{\mu\beta} P^{\nu\alpha} \over 2 }-{P^{\mu\nu}P^{\alpha\beta}\over 3} \right)\left(\partial_\alpha q_\beta\right)\,,\\
&\{\chi,\theta,\lambda\} ~\text{are arbitrary constants}\,.
\end{split}
\end{equation}
Note the above stress tensor has a structure very similar to that of BDNK stress tensor barring just the last term.
The dispersion polynomial has the following form
\begin{equation}\label{eq:discomp}
\begin{split}
P(\omega,k) &= \underbrace{\left(\omega +i\lambda~ k^2\right)}_{{\cal F}_1}\underbrace{\left( 3\omega^2- k^2 + 4 i \lambda~\omega k^2\right) }_{{\cal F}_2}\underbrace{(1 - i a_1~\omega)}_{{\cal F}_3}\underbrace{\left(
\left(1-i\omega \,a_1 \right) \left(1-i\omega \, b_1\right)+ a_2 b_2 \frac{k^2}{3}\right)}_{{\cal F}_4}.
\end{split}
\end{equation}
Here the zeroes ${\cal F}_1$ and ${\cal F}_2$ generate the hydrodynamic modes whereas the zeroes of the factors ${\cal F}_3$ and ${\cal F}_4$ generate non-hydrodynamic modes. ${\cal F}_3$ corresponds to the shear channel and ${\cal F}_4$ corresponds to the sound channel. Comparing ${\cal F}_3$ and ${\cal F}_4$ with the expression of ${\cal F}$ as given in equation \eqref{eq:quote}  we could easily read off a set of consistent solutions for the functions $R_T$, $R_u$, $F_u$, $F_T$, $G_u$ and $G_T$
\begin{equation}
    F_u = a_1\,,\quad F_T = {\frac{a_2}{\sqrt{3}}}\,,\quad G_T = b_1\,, \quad G_u = \frac{b_2}{\sqrt{3}}\,, \quad R_T=R_u=0\,.
\end{equation}
It is easy to see that the linearized equations of motion can be expressed in matrix form as
\begin{equation}\label{psdolin}
\begin{split}
4T^4\begin{bmatrix}
  -3i\omega&ik &0\\
  ik &-i\omega+ {4\over 3}k^2\lambda&0\\
  0&0&k^2\lambda-i\omega 
   \end{bmatrix}\cdot\underbrace{\begin{bmatrix}
 1 -i\omega~b_1&i\frac{k}{\sqrt{3}} ~b_2&0\\
  i \frac{k}{\sqrt{3}}~ a_2 &1-i\omega ~a_1&0\\
  0&0&1-i\omega ~a_1
   \end{bmatrix}}_{{\bf 1} +S}
    \begin{bmatrix}
    \delta T\\
    \beta_k\\
    \beta_\perp
    \end{bmatrix} =0 \,.
\end{split}
\end{equation}
Inverse frame transformation in this case will be generated by the inverse of the matrix $\left[{\bf 1} +S\right]$, which is clearly going to be an infinite series if expanded in powers of $\omega$ and $k$, indicating an infinite order frame transformation. The case discussed in the previous section is just a special case where $a_1 =c,~a_2=b_1=b_2 =0$.

Now, let us contrast this analysis with that of the BDNK theory. In the stress tensor described in equation \eqref{pseudo2}, if we choose $a_1 =a_2 \equiv a$ and $b_1 =b_2 \equiv b$,
and also set the last term (proportional to $\tilde\sigma^{\mu\nu}$) to zero, then the stress tensor becomes that of the simplest  BDNK theory. However, this simple modification (particularly the one involving the removal of the last term proportional to $\tilde\sigma_{\mu\nu}$) leads to a major change in the dispersion polynomial
\begin{equation}
\begin{split}
&P_\text{BDNK}(\omega,k) =
 \underbrace{\left[\omega  (1- ia~ \omega )+i\lambda k^2  \right]}_{{\cal G}_1}\\
 &\times \underbrace{\bigg\{\left[3 \omega 
   (1-ib~ \omega)-ia~ k^2\right] \left[3\omega  (1-ia~ \omega )-i(b-4\lambda)~ k^2\right]+3i k^2 [1-i(a+b)~\omega ]^2 \bigg\}}_{{\cal G}_2}.
\end{split}
\end{equation}
We observe that the non-hydrodynamic modes no longer factorize as they do in equation \eqref{eq:discomp} through factors ${\cal F}_3$ and ${\cal F}_4$. In $P_\text{BDNK}$, the analytic factors ${\cal G}_1$ and ${\cal G}_2$ encompass both hydrodynamic and non-hydrodynamic modes. This is a clear indication that the non-hydrodynamic modes in the BDNK theory are not simply artefacts of frame redefinition but are indeed physical.
\section{Summary and Outlook}
\label{sec:conclude}
The benchmark criteria for a hydrodynamic theory (constructed in terms of systematic derivatives corrections away from equilibrium) to be physically acceptable are causal wave propagation and stability against small perturbations. 
When we mention ``physical constraints'' in the article, we refer to these two criteria to validate a given theory. 
Now it is known that the non-hydrodynamic modes of the theory (which also sets the scale for a given theory) decide these criteria~\cite{Pu:2009fj,Kovtun:2019hdm} and the associated allowed range of the transport coefficients sets the parameter space for the theory where it is both stable and causal, and hence, physically acceptable for practical applications. Now, since we know that across a hydrodynamic frame transformation, additional non-hydrodynamic modes can pop up~\cite{Bhattacharyya:2023srn}, therefore, those modes may generically seem to impose further constraints for the theory to be causal and stable. But, since the fundamental properties of a theory, such as causality and stability, are not supposed to be frame-dependent, we must not trust these new constraints to validate the theory considering these new non-hydrodynamic modes to be purely ``frame-artefacts'' and ``not physical''. 
This work provides a novel way to segregate between such ``physical'' and ``artifact'' non-hydrodynamic modes such that the ``violation of physical constraints'' (stability-causality) by these ``artefact'' non-hydrodynamic modes should not discredit the physical acceptability of a theory.

One of the main conclusions of our paper is: suppose we start with a given energy-momentum tensor and then do an all-order field redefinition of the fluid variables that removes the non-hydrodynamic modes present in the original stress-energy tensor. The resulting stress-energy tensor (after field redefinition) will tell us whether the original non-hydrodynamic modes are to be trusted to provide physical benchmark for the theory. The following two scenarios can arise:
\begin{itemize}
\item After the field redefinition, if the modified stress-energy tensor (with no non-hydrodynamic modes) has an infinite number of terms involving arbitrarily high-order derivatives of the fluid variables, then the non-hydrodynamic modes present in the original stress-energy tensor are physical. 
\item However, after the field redefinition, if the modified stress-energy tensor is finite, then it means that the information about the non-hydrodynamic mode can simply be erased by field redefinition. This indicates that the non-hydrodynamic modes present in the original stress-energy tensor are mere artifacts or unphysical. Hence, even if these non-hydrodynamic modes violate physical constraints, they do not have any physical significance.
\end{itemize}
In short, after the infinite-order field redefinition, if the modified stress-energy tensor (with no non-hydrodynamic modes) has an infinite number of terms, then the non-hydrodynamic modes present in the original theory are physical. If the modified stress-energy tensor is finite, then those non-hydrodynamic modes are not physical.

In this work, we have demonstrated that in relativistic fluid dynamics, the spectrum of hydrodynamic modes remains unaffected by any redefinition of fluid fields. Conversely, the spectrum of non-hydrodynamic modes is subject to change under such redefinitions. In fact,
it is possible to entirely eliminate a non-hydrodynamic mode from the spectrum through an appropriate all-order frame redefinition. We have illustrated this phenomenon using the example of the first-order BDNK theory. However, such a field redefinition cannot completely erase the information of any physical non-hydrodynamic mode.
In the space of complex frequency and momentum, the infinite series of the hydrodynamic stress tensor will have a radius of convergence at the location of the lowest non-hydrodynamic mode~\cite{Heller:2013fn,Heller:2020uuy}. This observation also suggests that the lowest non-hydrodynamic mode in the theory acts as a cutoff in the hydrodynamic expansion~\cite{Grozdanov:2019kge,Heller:2020hnq,Cartwright:2021qpp}.

Our analysis also implies that in a causal physical theory, the removal of physical non-hydrodynamic modes necessitates the inclusion of all higher-order terms in the energy-momentum tensor without any truncation, and in Fourier space, the information
about the physical non-hydrodynamic mode will be embedded in the radius of convergence
of this infinite series~\cite{Bemfica:2019knx,Heller:2022ejw,Gavassino:2021owo,Gavassino:2023myj,Gavassino:2024pgl,Wang:2023csj}. 
If the energy-momentum tensor is truncated to any finite order, the presence of physical non-hydrodynamic modes becomes necessary to maintain causality. 
It has been observed, recently, that if the physical theory does not have non-hydrodynamic mode and is truncated to finite order, then it will not be causal, however, it may be stable~\cite{Hiscock:1985zz,Novak:2019wqg,deBoer:2020xlc,Armas:2020mpr,Basar:2024qxd,Pu:2009fj}. 
This is an interesting aspect which we leave for future work to prove for the general case.

We have also demonstrated that in certain special cases, a non-hydrodynamic mode may be merely an artefact of the chosen fluid fields. In such instances, it is indeed possible to completely remove the information of this mode from the spectrum. When this occurs, the dispersion polynomial will neatly factorize between the physical and artefact modes, resulting in a very precise cancellation after the frame transformation, leading to a finite number of terms in the stress tensor (assuming the stress tensor initially contained a finite number of terms). We have illustrated this with a simple example. It is important to note that these artefact modes do not impact the validity of the underlying hydrodynamic theory, even if they do not meet physical criteria such as stability and causality.

Finally, in the last section, we have demonstrated that adding a `specific' second-order term to the BDNK stress tensor results into clean hydrodynamic and non-hydrodynamic mode separation. The corresponding ``hydro-frame'' transformation can render it equivalent to the first-order relativistic Navier-Stokes stress tensor which is known to have finite number of terms with no non-hydrodynamic modes. Hence, the non-hydrodynamic modes of this `special' second order BDNK stress tensor are artefacts. 
It means that addition of this specific second-order term has converted the genuine non-hydrodynamic modes of the original first-order BDNK theory into frame artefacts.
In other words, when refining the BDNK theory with second-order corrections, one must be cautious to avoid such terms, as outlined in equation \eqref{pseudo2}. 

Our analysis has focused on the simplest case of uncharged fluids, where the conservation of the stress tensor serves as the sole hydrodynamic equation of motion. It may be of interest to extend this analysis to other types of fluids that satisfy various charge conservation equations, as well as to MIS-type theories, which involve auxiliary fields such as the shear tensor as fluid variable that could also be redefined in principle.

We have observed that a typical frame redefinition at the linearized level generates new non-hydrodynamic modes. These modes correspond to the zero eigen-space of the linearized frame transformation. In other words, if a perturbation initially has frequencies matching these new modes, it vanishes after the frame transformation, thereby trivially satisfying the linearized equation of motion. However, in the transformed frame, these would appear as new modes that alter the UV behavior of the theory without impacting its low-energy physics. This characteristic of frame redefinition can be utilized in various ways. In this work, we have employed it to construct an equivalent hydrodynamic theory with the non-hydrodynamic modes removed from the spectrum of linearized perturbations. Another potential application is to regulate the UV behavior of the theory, making the equations of motion compatible with computational requirements without altering the underlying physics. It would be interesting to explore the impact of such field redefinitions in the context of gravitational theories and their EFT corrections. For instance, in~\cite{Figueras:2024bba}, the authors used this technique in higher-derivative gravity theories to convert the equations into a well-posed initial value problem. Investigating a similar line in the context of quasi-normal modes of black branes and their fluid duals might provide us insights of using holography to write stable-causal hydrodynamic theories.
\begin{acknowledgments}
S.B., S.M. and S.R. acknowledge the Department of Atomic Energy, India, for the funding support.
R.S. acknowledges kind hospitality and support of INFN Firenze, Department of Physics and Astronomy, University of Florence, ECT* Trento, NCBJ Warsaw, TU Darmstadt, ITP Goethe University where part of this work is completed and a postdoctoral fellowship of West University of Timișoara, Romania. We acknowledge enlightening discussions with L.~Gavassino, M.~Spalinski, and A.~Yarom.
\end{acknowledgments}
\bibliography{pv_ref}{}
\bibliographystyle{utphys}
\end{document}